\documentclass[lettersize, journal]{IEEEtran}
\usepackage{amsmath, amsfonts}
\usepackage{algorithmic}
\usepackage{algorithm}
\usepackage{array}
\usepackage[caption=false, font=normalsize, labelfont=sf, textfont=sf]{subfig}
\usepackage{textcomp}
\usepackage{stfloats}
\usepackage{url}
\usepackage{verbatim}
\usepackage{graphicx}
\usepackage[table]{xcolor}
\usepackage{cite}
\usepackage{booktabs}
\usepackage{longtable}
\usepackage{ragged2e} % For \RaggedRight

\usepackage{makecell} % For \\ in table cells, useful for complex text
\usepackage{lipsum} % For dummy text to demonstrate page breaks

\usepackage{color, bbding, multicol, multirow}
\def \eg {\emph{e.g.}}
\def \ie {\emph{i.e.}}
\def \etal {\emph{et al.} }

\newcommand\red[1]{\textcolor{red}{#1}}

\newcommand\black[1]{\textcolor{black}{#1}}

\usepackage{pifont}
\begin{document}

\title{PEMUTA: Pedagogically-Enriched Multi-Granular Undergraduate Thesis Assessment}

\author{
    Jialu Zhang\IEEEauthorrefmark{1}, 
    Qingyang Sun\IEEEauthorrefmark{1}, 
    Qianyi Wang, 
    Weiyi Zhang, 
    Zunjie Xiao, 
    Xiaoqing Zhang\IEEEauthorrefmark{2}, 
    Jianfeng Ren, ~\IEEEmembership{Senior Member, ~IEEE, } 
    and Jiang Liu\IEEEauthorrefmark{2}, ~\IEEEmembership{Senior Member, ~IEEE}
%         % <-this % stops a space
        
\thanks{J. Zhang, Q. Sun, Q. Wang, W. Zhang, Z. Xiao, and X. Zhang are with the Research Institute of Trustworthy Autonomous Systems and Department of Computer Science and Engineering, Southern University of Science and Technology, Guangdong 518055, China (email: zhangjl3@sustech.edu.cn; 12332448@mail.sustech.edu.cn; 12111003@mail.sustech.edu.cn; 12231145@mail.sustech.edu.cn; zhangwy2022@mail.sustech.edu.cn; xq.zhang2@siat.ac.cn).}

\thanks{J. Ren, is with the School of Computer Science, University of Nottingham Ningbo China, Zhejiang 315100, China. (e-mail: jianfeng.ren@nottingham.edu.cn).}

\thanks{J. Liu is with the Research Institute of Trustworthy Autonomous Systems and Department of Computer Science and Engineering, Guangdong 518055, China, and also with the School of Computer Science, University of Nottingham Ningbo China, Zhejiang 315100, China, and also with the School of Ophthalmology and Optometry, Wenzhou Medical University, Zhejiang 325035, China (e-mail: liuj@sustech.edu.cn).}

\thanks{\IEEEauthorrefmark{1}These authors contributed equally to this work.}
\thanks{\IEEEauthorrefmark{2}These authors are co-corresponding authors for this work.}
}

% % The paper headers
% \markboth{Journal of \LaTeX\ Class Files, ~Vol.~14, No.~8, August~2021}%
% {Shell \MakeLowercase{\textit{\etal}}: A Sample Article Using IEEEtran.cls for IEEE Journals}

% \IEEEpubid{0000--0000/00\$00.00~\copyright~2021 IEEE}
% Remember, if you use this you must call \IEEEpubidadjcol in the second
% column for its text to clear the IEEEpubid mark.

\maketitle

\begin{abstract}
The undergraduate thesis (UGTE) plays an indispensable role in assessing a student's cumulative academic development throughout their college years. Although large language models (LLMs) have advanced education intelligence, they typically focus on holistic assessment with only one single evaluation score, but ignore the intricate nuances across multifaceted criteria, limiting their ability to reflect structural criteria, pedagogical objectives, and diverse academic competencies. Meanwhile, pedagogical theories have long informed manual UGTE evaluation through multi-dimensional assessment of cognitive development, disciplinary thinking, and academic performance, yet remain underutilized in automated settings. Motivated by the research gap, we pioneer PEMUTA, a pedagogically-enriched framework that effectively activates domain-specific knowledge from LLMs for multi-granular UGTE assessment. Guided by Vygotsky's theory and Bloom's Taxonomy, PEMUTA incorporates a hierarchical prompting scheme that evaluates UGTEs across six fine-grained dimensions: Structure, Logic, Originality, Writing, Proficiency, and Rigor (SLOWPR), followed by holistic synthesis.
Two in-context learning techniques, \ie, few-shot prompting and role-play prompting, are also incorporated to further enhance alignment with expert judgments without fine-tuning. We curate a dataset of authentic UGTEs with expert-provided SLOWPR-aligned annotations to support multi-granular UGTE assessment. Extensive experiments demonstrate that PEMUTA achieves strong alignment with expert evaluations, and exhibits strong potential for fine-grained, pedagogically-informed UGTE evaluations.

\end{abstract}
  
\begin{IEEEkeywords}
Multi-Granular Undergraduate Thesis Assessment, Education Agent, Large Language Models, Vygotsky's Theory, Bloom's Taxonomy, Education Intelligence.
\end{IEEEkeywords}

\section{Introduction}

\begin{figure}[t]
    \centering
    \includegraphics[width=\linewidth]{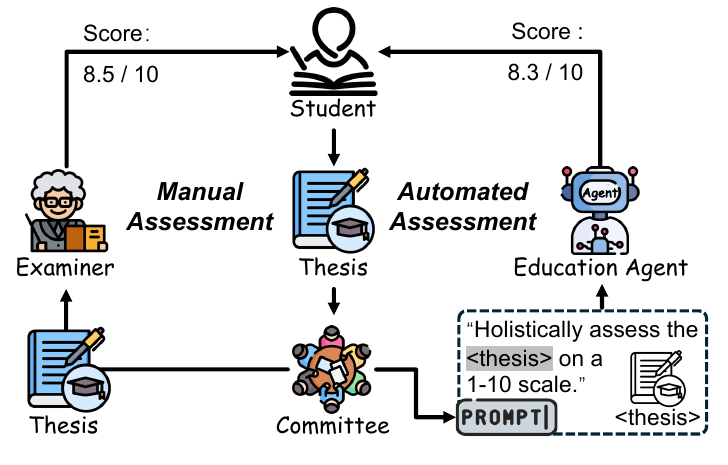}
    \caption{Comparison of manual and automated methods for UGTE assessment. Manual assessment, constrained by time and workload, typically yields a single holistic score. Current education agents improve scalability and consistency but still rely on generic prompts and holistic evaluations, lacking criterion-specific feedback.}
    \label{fig:intro}
\end{figure}

\IEEEPARstart{U}{dergraduate} thesis (UGTE) is a significant and comprehensive means for undergraduate education quality assessment, involving graduation eligibility and degree classification. The key role of UGTE is to evaluate how well students can apply disciplinary knowledge and academic competencies (\eg, analytical reasoning, research rigor, and intellectual originality) to solve real-world applications or theoretical problems~\cite{Chan_2001_FYP}. UGTEs are typically highly structured and substantially lengthy, often exceeding 50 pages, and must be evaluated across multifaceted criteria such as structural coherence, academic rigor, and methodological soundness. Consequently, manual UGTE assessment is inherently time-consuming and labor-intensive, posing longstanding challenges for educators/professors~\cite{LIN2023101830,Arias_2014_MasterThesis}. These challenges are further exacerbated by rising submission volumes and limited faculty resources, hindering the delivery of effective, timely, consistent, and personalized feedback. Developing assisted UGTE assessment approaches has garnered increasing attention from a broader perspective ~\cite{Xiao_2025_AESACL, Song_2024_AESAER}.

Early rule-based methods~\cite{Attali_2006_ERater, phandi_2015_AESLinear, taghipour2016neural, dong2016automatic} and recent deep neural networks~\cite{Uto_2023_HybridAES, Wang_2022_multiBert, zhang2021multimedia} have promoted the development of automated essay scoring (AES), but they are ill-suited for UGTE assessment due to their limited capacity to capture high-level semantics, and dependence on costly, large-scale annotated datasets. More recently, educational agents powered by large generative language models (LLMs) have advanced the development of Education Intelligence (EI) and gained increasing attention~\cite{Kostic_2024_LLMEI, Zhang_2025_LLMEdu, Lang_2025_GAIEI,imed_2025_EIDefinition,Qingyang_2025_EI_CPR}. LLMs are pre-trained on massive and diverse corpora, these models encode rich linguistic patterns and domain knowledge, enabling strong contextual reasoning and long-form text comprehension. Crucially, they can be adapted to complex educational tasks via prompt-based instructions without requiring additional fine-tuning~\cite{Xiao_2025_AESACL}.

As illustrated in Fig.~\ref{fig:intro}, LLMs can be specifically adapted as UGTE assessment agents, receiving assessment instructions along with the thesis text and returning a single final score. While this paradigm offers scalability and efficiency, three key limitations still exist. First, most existing systems rely on holistic scoring, providing only a single coarse-grained score per UGTE submission. \textbf{This ignores the intricate nuances among different academic competencies} such as analytical reasoning, research rigor, and intellectual originality, thereby limiting the interpretability of the results and the provision of criterion-referenced, actionable feedback~\cite{he_2022_AESwithMultipleTraits, Lagakis_2021_AESReview}. Second, \black{existing prompt methods often lack domain-specific or pedagogical grounding}, \textbf{failing to activate LLMs' internal representations of academic standards or align with established educational frameworks} like Bloom's Taxonomy and Vygotsky's theory. As a result, they often neglect critical indicators close to UGTE assessment, such as intellectual development, and yield context-insensitive feedback with limited diagnostic and instructional value~\cite{Dimitrova_2011_ScaffoldingDT, Goh_2011_ThesisBloom, Song_2024_AESAER}. \black{Third, UGTE evaluation is inherently multi-dimensional, \textbf{requiring complex reasoning that generic or monolithic prompts fail to elicit effectively}}~\cite{haagsman_2021_ThesisRubric, Cohn_2024_CoTEduLLM}.

To tackle these challenges, we propose the \textbf{\underline{P}}edagogically-\textbf{\underline{E}}nriched \textbf{\underline{M}}ulti-granular \textbf{\underline{U}}ndergraduate \textbf{\underline{T}}hesis \textbf{\underline{A}}ssessment (\textbf{PEMUTA}), a novel framework that guides LLMs in mining domain-specific knowledge to deliver both criterion-level evaluations and synthesized holistic judgments. However, there remains a lack of standardized fine-grained evaluation dimensions tailored specifically for UGTE. To fill this gap, we draw inspiration from two foundational pedagogical theories widely adopted in manual UGTE assessment: Vygotsky's theory and Bloom's Taxonomy~\cite{Goh_2011_ThesisBloom, Dimitrova_2011_ScaffoldingDT}. Vygotsky's theory emphasizes the developmental and epistemological aspects of UGTE, and Bloom's Taxonomy offers a structured hierarchy of cognitive skills used in instructional design and assessment. Building on these insights, we define six pedagogically grounded dimensions: \textbf{\underline{S}}tructure, \textbf{\underline{L}}ogic, \textbf{\underline{O}}riginality, \textbf{\underline{W}}riting, \textbf{\underline{P}}roficiency, and \textbf{\underline{R}}igor, collectively abbreviated as \textbf{SLOWPR}. LLMs are prompted to evaluate UGTEs across multiple levels of granularity, leveraging their rich textual priors to produce criterion-aligned, multi-granular feedback. By integrating both cognitive and developmental perspectives, PEMUTA improves alignment with expert evaluations while enhancing interpretability, pedagogical relevance, and diagnostic value.

Additionally, the multi-granular assessment typically involves processing rich, dimension-specific instructions, which poses challenges to the direct application of LLMs~\cite{Shi_2023_LLMDistraction,Bizhu_2025_CVPR,zhang_2024_LLMDistraction}. The extensive information embedded within detailed prompts may lead to attention distraction across subtasks, misalignment of criteria-specific knowledge across dimensions, and ultimately reduce overall coherence. To address this issue, we propose a hierarchical prompting strategy that guides the model through a structured two-stage evaluation process: first, performing dimension-level assessments, then synthesizing a holistic judgment. This design improves alignment with human evaluators while enhancing interpretability, consistency, and pedagogical relevance.
To further improve alignment, we incorporate two in-context learning techniques: few-shot prompting, which supplies exemplars structured according to the proposed multi-granular framework, and role-play prompting, which conditions the model to act as an experienced academic evaluator. These techniques together enhance assessment performance without additional fine-tuning. Extensive experiments demonstrate that PEMUTA achieves strong agreement with expert ratings across both fine-grained and holistic UGTE evaluation.

Our contributions can be summarized as follows. 
1)~We rethink automated UGTE assessment and pioneer the PEMUTA, a pedagogically-enriched hierarchical prompting framework for structured evaluations across varying levels of granularity.
2)~We integrate Vygotsky's theory and Bloom's Taxonomy into the assessment process, which guides the LLM to evaluate UGTEs across six critical SLOWPR dimensions, eliciting rich academic priors to produce criterion-based, pedagogically-grounded evaluations that reflect both academic performance and developmental potential. 
3)~We develop an In-context Enhanced Hierarchical Prompting scheme that incorporates few-shot and role-play prompting to enhance alignment with expert judgment without model fine-tuning. It decomposes multi-granular UGTE assessment into dimension-specific evaluations followed by holistic synthesis, effectively activating criteria-specific reasoning and enabling interpretable, multi-level evaluation. 
4)~We curate the MUTA dataset to support comprehensive UGTE evaluation. Experiments show that PEMUTA achieves strong agreement with human experts on both traditional holistic and novel fine-grained dimensions, demonstrating its effectiveness and robustness.

\section{Related Work}
\subsection{Automated Essay Scoring}
Automated Essay Scoring (AES) refers to the methods that automatically evaluate and assign scores to student-written essays in educational contexts, aiming to provide scoring that is consistent, valid, and efficient, offering performance comparable to that of human raters~\cite{ke_2019_AESSurvey,Xiao_2025_AESACL,Song_2024_AESAER}. AES systems typically treat the task as a classification problem, wherein essays are mapped to predefined score levels~\cite{ke_2019_AESSurvey,Lagakis_2021_AESReview,phandi_2015_AESLinear}. Early AES systems predominantly relied on heuristic or rule-based approaches and manually engineered features. These systems applied statistical learning algorithms such as linear regression, support vector machines (SVMs), decision trees, and ensemble models to estimate overall scores through a weighted combination of sub-scores representing aspects such as grammar, vocabulary, organization, and coherence~\cite{ke_2019_AESSurvey,Lagakis_2021_AESReview}. E-rater, developed by the Educational Testing Service (ETS), is one of the most influential early systems. It incorporates approximately fifty predefined linguistic features derived from syntax, discourse structure, and lexical usage, combined with regression-based models for score prediction~\cite{Attali_2006_ERater}. Another widely used system, IntelliMetric by Vantage Learning, utilizes over four hundred features spanning semantic, syntactic, and rhetorical levels. It processes input through multi-stage pipelines that include hierarchical linguistic parsing and statistical modeling~\cite{Rudner_2006_IntelliMetric}. These systems benefit from interpretability, since their predictions are based on explicit and well-defined features. However, they also face significant limitations. Hand-crafted features require extensive domain expertise and manual effort to design and adapt. Moreover, rule-based models often fail to capture deeper semantics, such as logical coherence, argument structure, and contextual relevance~\cite{Wang_2022_multiBert,li_2024_AES}. These limitations have motivated a shift towards data-driven approaches that automatically learn high-level representations from large-scale corpora~\cite{he2022automated_trait,dong2016automatic,faseeh2024hybrid}.

Deep learning has significantly advanced AES~\cite{Alikaniotis2016AES,faseeh2024hybrid,li_2024_AES}. Early models leveraged recurrent and convolutional networks (CNN) to learn semantic and syntactic representations from raw text.
Taghipour and Ng applied convolutional layers over word embeddings, followed by a bi-directional LSTM to jointly capture local n‑gram patterns and long-range dependencies~\cite{taghipour2016neural}. 
Alikaniotis \etal introduced score-specific word embeddings and used a bi-directional LSTM to model contextual dependencies in essays while emphasizing correlations between word usage and scoring outcomes~\cite{Alikaniotis2016AES}.
  Dong and Zhang proposed a hierarchical two-layer CNNs to extract multi-scale abstractions at both word and sentence levels~\cite{dong2016automatic}. 
  More recently, transformer-based models and attention mechanisms have achieved further improvements in AES by capturing global context and modeling long-distance dependencies~\cite{faseeh2024hybrid,li_2024_AES}. 
  Wang \etal, proposed a multi-scale BERT framework that jointly learns token-, segment-, and document-level representations for essay evaluation~\cite{Wang_2022_multiBert}. 
  Faseeh \etal introduced a hybrid model that integrates handcrafted linguistic features with contextualized RoBERTa embeddings. The combined features are passed to a lightweight XGBoost regressor to enhance both predictive performance and interpretability~\cite{faseeh2024hybrid}.
  Uto \etal, incorporated Item Response Theory and proposed a hybrid framework that aggregates outputs from multiple AES models via the Generalized Many-Facet Rasch Model (GMFRM)~\cite{Uto_2023_HybridAES}. 
  While these advances offer improved accuracy and flexibility, they often require complex model architectures, high computational resources, and large-scale annotated datasets. Such requirements limit their scalability and can lead to overfitting and poor generalization when applied in real-world educational settings~\cite{li_2024_AES}.

Traditional AES methods have demonstrated notable progress, but they heavily rely on manual feature engineering or task-specific supervision, reducing their adaptability to complex undergraduate thesis assessment. Recent advances in large language models present a promising alternative by enabling flexible, prompt-based assessment paradigms. 
  
\subsection{LLMs in Automated Assessments}

The emergence of generative large language models (LLMs) has introduced a new paradigm in educational assessment~\cite{Qingyang_2025_EI_CPR,Zhang_2025_LLMEdu}. Pre-trained on massive text corpora, these models encode rich linguistic knowledge and demonstrate strong capabilities in capturing deep semantics, discourse structure, and contextual coherence, all of which are critical for effective essay evaluation~\cite{Song_2024_AESAER}. Without requiring additional fine-tuning, LLMs can be adapted to essay scoring via prompt engineering. Typical prompting strategies include (1) few-shot prompting, which provides task instructions along with a few labeled exemplars, and (2) zero-shot prompting, which uses only task instructions. These approaches significantly reduce the need for labeled datasets and hand-crafted features that traditional AES systems depend on~\cite{Song_2024_AESAER,Kostic_2024_LLMEI,Xiao_2025_AESACL}.

However, parameter-free LLM-based AES systems face several critical limitations. Most methods produce only a single holistic score, resulting in limited interpretability and minimal pedagogical value~\cite{Maier_2022_PersonalizedFeedback, suraworachet2023impact}. Haagsman \etal empirically investigated the limitations of holistic-only scoring for UGTE assessment. They constructed a rubric-based framework with thirteen categories, each defined by one to six sub-criteria. Human examiners were instructed to assign an overall grade and individual scores for each criterion on a five-point scale ranging from insufficient to good~\cite{haagsman_2021_ThesisRubric}. They found that UGTEs receiving identical overall scores often exhibited wide variations in rubric-level evaluations, highlighting the need for interpretable, multi-granular assessment systems for complex academic evaluation. While LLMs offer potential for fine-grained, rubric-aligned scoring, their effectiveness depends on precisely structured prompts. Without domain-specific cues, models may default to generic or irrelevant knowledge, yielding misaligned assessments~\cite{Qingyang_2025_EI_CPR,he_2022_AESwithMultipleTraits}. These problems are exacerbated in complex writing tasks such as thesis evaluation, which requires nuanced disciplinary knowledge and pedagogical alignment~\cite{Qingyang_2025_EI_CPR}. Hallucination and potential attention distraction commonly observed in LLMs further compromise output coherence and assessment reliability~\cite{Shi_2023_LLMDistraction,zhang_2024_LLMDistraction,Cohn_2024_CoTEduLLM}.

To address these challenges, recent studies have proposed several strategies focused on managing prompt structure~\cite{tang_2025_PromptCompression,Wang_2024_TaskSpecificPrompt,mu_2023_PromptCompression}. Some researchers modularize detailed prompts into segments processed sequentially or conditionally, enabling the model to maintain focus on localized information before integrating global understanding~\cite{zhang_2024_LLMDistraction,Cohn_2024_CoTEduLLM, Liu_2023_HierarchicalPrompt}. Liu \etal, introduced a hierarchical prompt learning strategy (HiPro), which quantifies inter-task affinity and constructs a task tree to capture task cross-granularity dependencies~\cite{Liu_2023_HierarchicalPrompt}. Cohn \etal further enhanced prompt design by integrating human-in-the-loop mechanisms. They combined few-shot and active learning techniques with chain-of-thought prompting to improve explanation quality in formative K–12 science assessments~\cite{Cohn_2024_CoTEduLLM}. Together, these approaches underscore the importance of structured and context-sensitive prompt design in reducing distraction and improving the reliability, focus, and interpretability of LLM-based assessments.

\section{Proposed Methodology}
\subsection{Overview of Proposed Method} 
\begin{figure*}[thpb]
  \centering
  \includegraphics[width=1\linewidth]{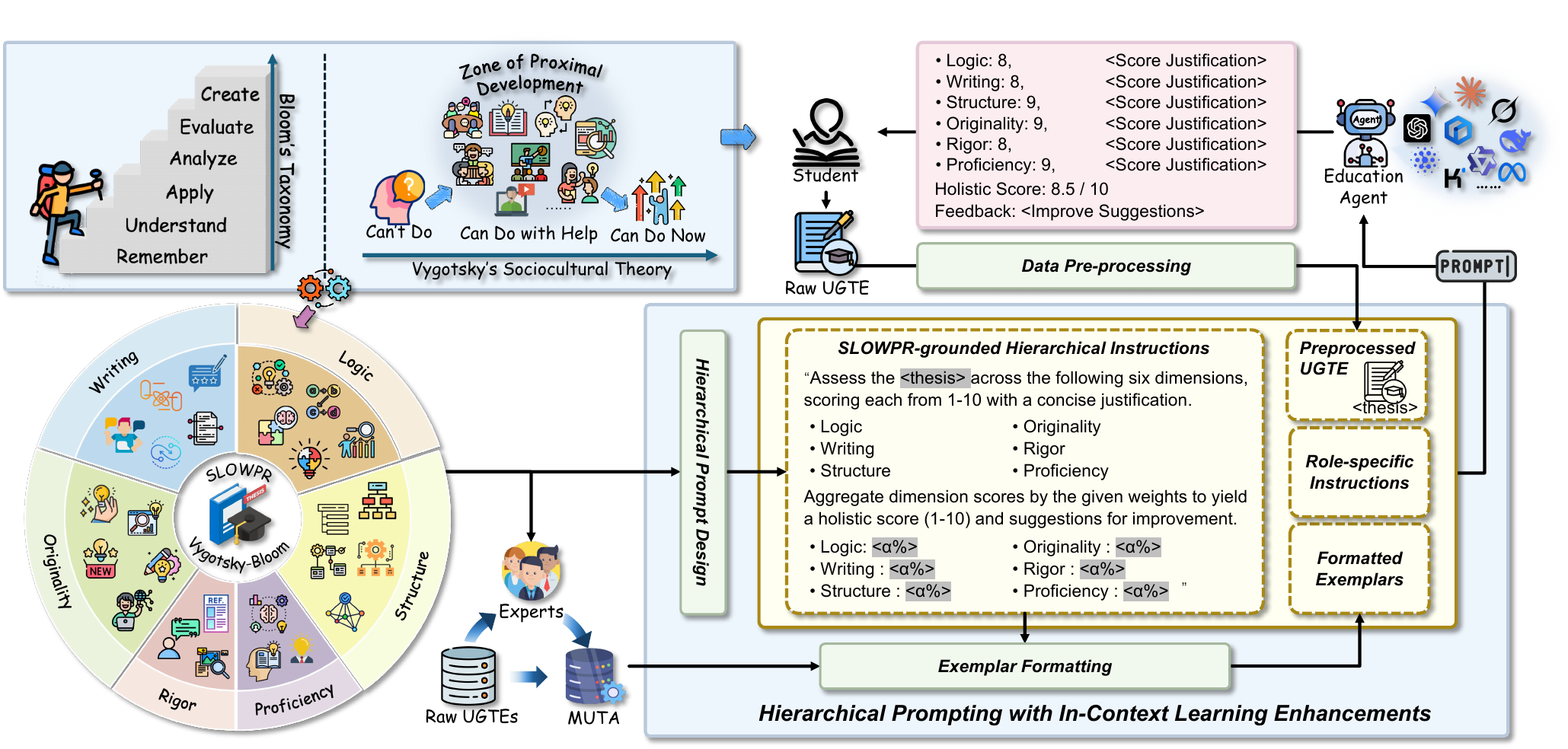}
  \caption{Overview of the proposed PEMUTA framework for undergraduate thesis assessment across multiple levels of granularity.}
  \label{fig:overview}
\end{figure*}

This paper proposes \textbf{PEMUTA}, a pedagogically-enriched framework for multi-granular assessment of undergraduate theses. The system architecture is illustrated in Fig.~\ref{fig:overview}. 
To support multi-granular evaluation, our proposed framework begins by defining six fine-grained assessment dimensions: \textbf{\underline{S}}tructure, \textbf{\underline{L}}ogic, \textbf{\underline{O}}riginality, \textbf{\underline{W}}riting, \textbf{\underline{P}}roficiency, and \textbf{\underline{R}}igor (\textbf{SLOWPR}), derived from the integration of Bloom's Taxonomy and Vygotsky's sociocultural theories. These dimensions enable the assessment to capture both academic performance and developmental potential, in line with contemporary educational objectives in higher education. The theoretical basis and operational definitions of SLOWPR are elaborated in Section~\ref{sec:SLOWPR}. In parallel, raw UGTEs in PDF format are processed into model-compatible plain text via a dedicated pipeline comprising two sequential stages: layout-aware content extraction and semantic document reconstruction. The final output is a structured JSON representation that preserves both structural integrity and semantic coherence of the original UGTE. Implementation details are described in Section~\ref{sec:dataset}.

A hierarchical prompting is then introduced to guide the LLM through a two-stage evaluation process. In the first stage, the model performs dimension-specific assessments, generating criterion-aligned scores and justifications for each SLOWPR aspect. To improve alignment with expert evaluations, two in-context learning techniques are incorporated: (1) \textit{few-shot prompting}, which supplies formatted UGTE exemplars matching the expected multi-granular output structure without accompanying fine-grained justifications or holistic feedback; and (2) \textit{role-play prompting}, which conditions the model to act as a thesis committee member.

The final composite prompt consists of: (i) the preprocessed thesis, (ii) SLOWPR-grounded hierarchical instructions, (iii) few-shot formatted exemplars, and (iv) role-specific instructions. This prompt is subsequently submitted to a capable LLM for assessment. The resulting output is a standardized evaluation report comprising: (a) fine-grained scores with supporting justifications, and (b) a holistic score accompanied by revised suggestions. Overall, PEMUTA offers a scalable, interpretable, and pedagogically grounded solution for comprehensive UGTE assessment.

\subsection{Pedagogically-Grounded Fine-grained Dimensions}
\label{sec:SLOWPR}

Fine-grained evaluations guided by pedagogical theories are often implicitly embedded throughout the process~\cite{Goh_2011_ThesisBloom, haagsman_2021_ThesisRubric, Arias_2014_MasterThesis}, enabling human assessors to capture diverse competencies such as reasoning quality, writing proficiency, and methodological rigor. Existing LLM-based systems~\cite{Song_2024_AESAER,Xiao_2025_AESACL} offer scalable and efficient solutions, however, they predominantly rely on simple standard prompts such as \textit{`Holistically assess the given thesis on a 1-10 scale.'} to generate a numerical score. Their responses often diverge from human judgments. We attribute this performance gap to the underutilization of domain-specific knowledge priors and the lack of structured reasoning in LLMs. Additionally, the holistic assessment paradigm often fails to provide actionable feedback that supports student learning and improvement. To address these limitations, we pioneer a multi-granular UGTE framework that evaluates UGTEs across both fine-grained dimensions and the holistic outcomes. The definition of these fine-grained dimensions is informed by established pedagogical theories that reflect how students develop academic competence throughout their undergraduate education.

Vygotsky's sociocultural theory emphasizes the role of scaffolded learning in the transition from externally supported activities to independent academic competence~\cite{Dimitrova_2011_ScaffoldingDT}. Undergraduate students engage in various guided academic practices, such as supervised research, collaborative problem-solving, and formal academic writing, that help internalize scholarly norms and skills. These experiences contribute not only to the acquisition of disciplinary knowledge but also to the broader educational goal of preparing graduates for professional practice and lifelong learning. As the capstone of undergraduate education, the UGTE should therefore be assessed not only in terms of its immediate academic quality but also in terms of how it reflects a student's developmental progression toward scholarly independence.

Complementing this developmental view, Bloom's Taxonomy offers a structured hierarchy of cognitive abilities: Remember, Understand, Apply, Analyze, Evaluate,and Create, which has widely adopted in thesis assessment rubrics~\cite{Goh_2011_ThesisBloom}. However, its assumption of linear progression does not always align with real-world student performance, \eg, a student may exhibit sophisticated analytical reasoning while struggling with foundational application. Moreover, Bloom's framework pays little attention to sociocultural influences and individual learning trajectories. Consequently, UGTE evaluations based solely on Bloom's Taxonomy risk underrepresenting emerging capabilities and fail to recognize developmental efforts.

Table~\ref{tab:theoretical_synergy} summarizes the complementary strengths and limitations of Bloom's Taxonomy and Vygotsky's theory in the context of UGTE assessment. Synthesizing insights from both perspectives, we develop a pedagogically grounded fine-grained evaluation framework that captures both demonstrated academic performance and students' developmental potential from UGTE. This framework comprises six core evaluation dimensions, detailed as follows.
\begin{table}[t]
\caption{Comparison of Bloom's Taxonomy and Vygotsky's Theory in the Context of UGTE Assessment}
\centering
\resizebox{\linewidth}{!}{\begin{tabular}{>{\raggedright\arraybackslash}m{1.3cm} >{\raggedright\arraybackslash}m{2.8cm} >{\raggedright\arraybackslash}m{2.8cm}}
\toprule
\textbf{Theory} & \textbf{Strengths} & \textbf{Limitations} \\
\midrule
Bloom & Structured hierarchy of cognitive processes; Widely adopted in rubric design and educational assessment. & Presumes linear progression; Overlooks sociocultural or developmental variability.  \\
\midrule
Vygotsky & Highlights learning potential and scaffolded development; Captures dynamic progression over time. & Lacks formalized structures for assessment; Less directly applicable to rubric-based evaluation. \\
\bottomrule
\end{tabular}}
\label{tab:theoretical_synergy}
\end{table}

\begin{table*}[h]
\centering
\caption{Aspects and Theoretical Justification of the Fine-grained Evaluation Dimensions}
\resizebox{\textwidth}{!}{\begin{tabular}{>{\raggedright\arraybackslash}m{1.3cm} >{\raggedright\arraybackslash}m{2.6cm} >{\raggedright\arraybackslash}m{12.4cm}}
\toprule
\textbf{Dimension} & \textbf{Aspects Evaluated} & \textbf{Theoretical Justification (Motivation/Vygotsky/Bloom)} \\
\midrule
\textbf{Structure} & 
Organization of chapters; Coherence across sections; Smooth transitions. & 
\textbf{Motivation:} A clearly structured undergraduate thesis supports effective presentation of project objectives, methods, and outcomes, demonstrating the student's ability to manage complex information and follow academic conventions.
\newline
\textbf{Vygotsky:} Students internalize the competence to structure their thesis through observing faculty examples, receiving mentor feedback, participating in scholarly discussions, and scaffolded practice under FYP supervision.
\newline
\textbf{Bloom:} Students \textit{Understand} formal UGTE structures and \textit{Apply} them to organize their own work. \\

\midrule

\textbf{Logic} & 
Consistency among research questions, methodology, and conclusions; Clarity of reasoning and argument. & 
\textbf{Motivation:} Logical coherence ensures the core components of the UGTE, including research objectives, methodology, analysis, and conclusions, are clearly connected and aligned. Students should demonstrate the ability to build coherent and evidence-based arguments throughout the thesis. \newline
\textbf{Vygotsky:} Students develop logical reasoning for UGTE writing through sustained academic interactions such as supervision meetings, literature reviews, and peer feedback, progressing from guided support to independent argumentation.\newline
\textbf{Bloom:} Students \textit{Analyze} complex research problems, methods, and data, \textit{Evaluate} evidence validity and argument coherence, and ultimately \textit{Create} a logically consistent thesis. \\

\midrule

\textbf{Originality} & 
Original perspectives; Novel research questions; Theoretical or methodological innovation. &
\textbf{Motivation:} As a culminating academic project, the FYP reflects not only mastery of knowledge but also student's independent thinking capacity. Originality serves as a key indicator, requiring critical engagement and re-framing with established ideas through the student's intellectual perspective.\newline
\textbf{Vygotsky:}  Through socially mediated practices, \eg, engagement with supervisors, peers, and academic texts, students internalize disciplinary thinking patterns. This internalization supports independent reasoning and critical reinterpretation of knowledge, enabling novel contributions in the FYP.\newline
\textbf{Bloom:}  Students \textit{Analyze} complex ideas, \textit{Evaluate} existing alternatives, and \textit{Create} new interpretations, insights, frameworks or solutions.\\

\midrule

\textbf{Writing} & 
Clarity; Grammatical accuracy; Academic tone; Adherence to disciplinary conventions. & 
\textbf{Motivation:} The UGTE is a formal academic document whose quality depends not only on research content but also on clear, precise writing that meets academic standards, avoiding colloquial expressions, vagueness, and language errors. \newline
\textbf{Vygotsky:} Students' academic writing skills develop through supervisory feedback, peer review, sustained exposure to scholarly texts, progressing toward independent mastery within the ZPD.
 \newline
\textbf{Bloom:} Students must \textit{Remember} grammatical and structural rules, \textit{Understand} disciplinary writing conventions, and \textit{Apply} them effectively to produce a coherent, discipline-appropriate thesis.\\

\midrule

\textbf{Proficiency} & 
Application of course knowledge; Use of technical terminology; Problem-solving skills with disciplinary understanding; Use of field-specific tools; & 
\textbf{Motivation:} Undergraduate education aims to develop domain-specific expertise. As its capstone, the FYP should demonstrate whether students can apply accumulated knowledge to to real or research-based problems within their field.\newline
\textbf{Vygotsky:}  Through scaffolded experiences such as supervision, academic discussion, and guided use of disciplinary tools, students internalize knowledge and practices, enabling independent application in complex FYP tasks.\newline
\textbf{Bloom:} Students \textit{Understand} theoretical concepts, \textit{Apply} them in practice or research, and \textit{Analyze} domain-specific problems accurately.\\

\midrule

\textbf{Rigor} & 
Source reliability; Citation accuracy and format; Compliance with academic ethics. & 
\textbf{Motivation:} Undergraduate education is expected to cultivate students' awareness of academic standards and critically thinking in both scholarly and professional contexts. Beyond general writing tasks, the UGTE must strictly adhere to academic conventions, ensuring that the work is credible and ethically sound. \newline
\textbf{Vygotsky:} Through ongoing exposure to exemplary scholarly work, supervisor feedback, and institutional training, students internalize practices of source evaluation, proper citation, and ethical standards, ultimately regulating their behavior independently during the FYP process. \newline
\textbf{Bloom:} Students \textit{Remember} citation rules and ethical guidelines, \textit{Understand} their role in academic integrity, \textit{Apply} them throughout the research and writing process, and \textit{Evaluate} the reliability and relevance of sources and research practices used in the thesis.\\

\bottomrule
\end{tabular}
}
\label{tab:SLOWPR-design}
\end{table*}

\noindent\textbf{Structure}: Evaluates whether the thesis is well-organized, with a coherent and logical chapter structure. This dimension reflects the student's ability to manage academic structure and maintain conceptual clarity throughout the document.

\noindent\textbf{Logic}: Assesses the clarity and coherence of argumentation, including the alignment between research questions, methodology, evidence, and conclusions. It captures higher-order cognitive skills, \ie, \textit{Analyze}, \textit{Evaluate}, and \textit{Create} in Bloom's Taxonomy, and also acknowledges scaffolded reasoning and logical development as conceptualized in Vygotsky's Zone of Proximal Development (ZPD).

\noindent\textbf{Originality}: Evaluates the novelty and insightfulness of the thesis, particularly in the formulation of new perspectives or solutions. This dimension corresponds to the skills up to the \textit{Create} level of Bloom's Taxonomy, and accounts for emerging creative potential within the ZPD.

\noindent\textbf{Writing}: Examines linguistic clarity, grammatical accuracy, academic tone appropriateness, and adherence to disciplinary writing conventions. Extends beyond the surface-level correctness emphasized by Bloom's Taxonomy, it considers students' internalization of academic language acquisition through engagement with social interaction and instructional scaffolding, such as feedback from instructors or peer reviews.

\noindent\textbf{Proficiency}: Assesses students' mastery of disciplinary knowledge, particularly their understanding, application, and analysis of theoretical concepts and domain-specific methods in practical contexts. This dimension aligns primarily with the \textit{Understand}–\textit{Apply}–\textit{Analyze} levels of Bloom's Taxonomy and incorporates Vygotsky's notion of knowledge transfer as cognitive development across contexts.

\noindent\textbf{Rigor}: Measures adherence to academic conventions and standards of academic integrity, including citation accuracy, source reliability, and ethical compliance. It aligns with Bloom's \textit{Apply} level and reflects the internalization of scholarly behaviors through exposure to supervisory guidance and iterative academic practice, as emphasized in Vygotsky's theory.

These six dimensions constitute the fine-grained criteria for UGTE assessment, emphasizing both the cognitive sophistication of thesis content and the developmental trajectory demonstrated through students' academic writing. Among them, the first four, Structure, Logic, Originality, and Writing, serve as the core indicators of thesis quality, while the remaining two, Rigor and Proficiency, function as supporting indicators. The specific aspects evaluated and their theoretical justifications are summarized in Table~\ref{tab:SLOWPR-design}.

\subsection{Hierarchical Prompting with In-Context Learning Enhancements}
\label{sec:hierarchical_prompting}

Multi-granular assessment within a single input requires integrating diverse evaluative criteria and task‑specific instructions into information-dense prompts. However, such configurations often diffuse the model's attention across dimensions, impairing relevant knowledge activation and degrading output coherence~\cite{Shi_2023_LLMDistraction,Bizhu_2025_CVPR,zhang_2024_LLMDistraction}. To address these challenges, we propose a hierarchical two-stage prompting mechanism that structures the evaluation into focused, pedagogically aligned steps.

In the first stage, the model performs dimension-specific evaluations independently across the six SLOWPR criteria. For each dimension \(d\in\{S, L, O, W, P, R\}\), the model generates a tuple \((y_i^d, j_i^d)\), where \(y_i^d\) denotes the numerical score and \(j_i^d\) the corresponding criterion-specific justification for the \(i\)-th sample. This decomposition reduces cross-dimensional interference and facilitates more targeted activation of domain-relevant knowledge.

Once the six fine-grained assessments are obtained, the model proceeds to the second stage, synthesizing them into a coherent holistic evaluation. The overall score \(y_i^{H}\) is computed as a weighted sum:

\begin{equation}
y_{i}^{H} = \sum_{d \in \mathcal{D}} w^d \cdot y_i^d, \quad \mathcal{D} = \{S, L, O, W, P, R\},
\end{equation}
where \(w^d \in [0,1]\) represents the weight for each dimension, and \(\sum_{d} w^d = 1\). These weights can be configured to reflect disciplinary emphasis or institutional rubrics, offering flexibility across different academic contexts. The model also produces formative feedback by integrating prior dimension-level evaluations with holistic insights, offering practical suggestions for student improvement.

To strengthen alignment with expert judgment without model fine-tuning, we complement the SLOWPR-grounded hierarchical prompts with two in‑context learning techniques: few-shot prompting and role-play prompting. Specifically, we select representative UGTE samples and format them to match the expected multi-granular output structure, including six SLOWPR dimension scores and a holistic rating. We exclude textual rationales, as the ground-truth annotations consist solely of numerical grades. Rather than relying on direct instructions, these few-shot exemplars provide pattern-based guidance, serving as an explicit chain-of-thought (CoT) trigger that helps the model internalize the structure and format of rubric-aligned evaluations. 

In contrast, role-play prompting functions as an implicit CoT trigger, shaping model behavior through persona conditioning. Prior studies show that assigning a role-specific persona helps activate domain-relevant knowledge and structured reasoning patterns~\cite{Song_2024_AESAER}. In addition, this role conditioning narrows the model's activation space, further reducing ambiguity and encouraging coherent, expert-like responses. Accordingly, we instruct the model to assume the role of an experienced UGTE committee member, prompting it to adopt a formal academic tone, discipline‑appropriate vocabulary, and expert evaluative reasoning aligned with institutional expectations.

\section{Construction of MUTA Dataset}
\label{sec:dataset}
To support the multi-granular UGTE assessment task, we constructed the \textbf{\underline{M}}ulti-granular \textbf{\underline{U}}ndergraduate \textbf{\underline{T}}hesis \textbf{\underline{A}}ssessment (\textbf{MUTA}) dataset. This dataset comprises 60 authentic undergraduate theses collected from Computer Science undergraduates at Southern University of Science and Technology (SUSTech) during the 2023 and 2024 academic years. All research procedures were approved by the SUSTech Ethics Committee and conducted in accordance with institutional guidelines and the Declaration of Helsinki. Written informed consent was obtained from all participants. All personally identifiable information, including student names, ID numbers, and supervisor details, was rigorously anonymized. MUTA is available exclusively for non-commercial academic use.

\begin{figure}[!t]
    \centering
    \includegraphics[width=0.85\linewidth]{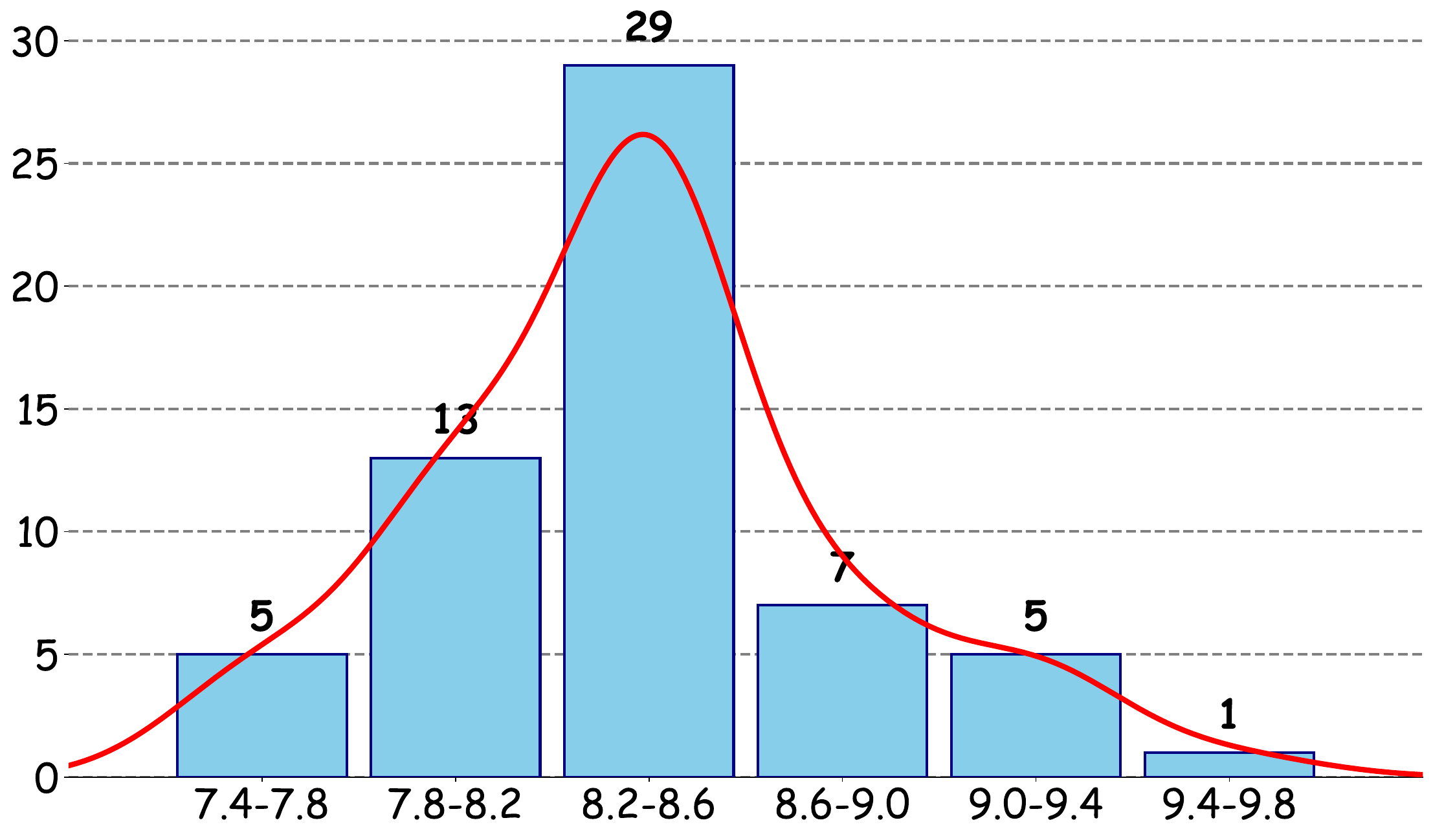}
    \caption{Histogram of the holistic score distribution of the collected dataset.}
    \label{fig:histogram}
\end{figure}

Each entry in MUTA includes the original thesis in PDF format, along with 0–10 scale ratings at multiple levels of granularity. The fine-grained scores are manually assigned by experienced evaluators based on the SLOWPR framework, while the holistic score is automatically computed as a weighted aggregation of the dimension-level ratings.
Fig.~\ref{fig:histogram} presents the distribution of holistic scores, with most falling in the 7.8-8.6 range. The 8.2-8.6 interval shows the highest density, accounting for $29$ UGTEs. Multi-granular scoring statistics are summarized in Table~\ref{tab:dataset2}. The mean holistic score is $8.4$ ($\sigma=0.43$), ranging from $7.4$ to $9.6$. Average scores across the six fine-grained dimensions are as follows: \textit{Logic} - $8.3$, \textit{Writing} - $8.2$, \textit{Structure} - $8.2$, \textit{Originality} - $8.4$, \textit{Proficiency} - $9.0$, \textit{Rigor} - $8.8$. Among these, $\textit{Rigor}$ exhibits the highest variability ($\sigma=1.53$), and $\textit{Logic}$ the lowest ($\sigma=0.46$).

\begin{table}[!t]
    \centering
    \caption{Statistics of expert-assigned scores across fine-grained dimensions and  overall holistic evaluation in the MUTA dataset. Std.\ denotes standard deviation.}
      \begin{tabular}{lcccr}
      \toprule
      Dimension & MEAN & Std. & Min & Max \\
      \midrule
      Logic & 8.31 & 0.46 & 7.5 & 9.5\\
      Writing & 8.15 & 0.58 & 7.0 & 9.0\\
      Structure & 8.20 & 0.68 & 6.5 & 10.0\\
      Originality & 8.36 & 0.72 & 7.0 & 9.5\\
      Proficiency & 9.05  & 0.50 & 8.0 & 10.0 \\
      Rigor & 8.83 & 1.54 & 3.0 & 10.0\\
      \midrule
      Holistic & 8.39 & 0.43 & 7.4 & 9.6 \\
      \bottomrule
      \end{tabular}
    \label{tab:dataset2}
\end{table}

The collected theses are originally stored in PDF format. However, raw PDF files cannot be directly processed by LLMs. Most LLMs require plain text encoded in UTF-8, with a linear structure. In contrast, PDFs are binary-encoded for visual presentation and human readability. Each page is rendered as a visual canvas, where text, images, tables, and other elements are placed using absolute coordinates. Although PDFs appear visually organized, they do not preserve the underlying linguistic and structural logic required for language model processing~\cite{Dagdelen2024}.

Extraction tools such as \textbf{\textit{pdfplumber}} often fail to recover this logic. In academic theses, repeated elements like headers, footers, and page numbers are frequently merged into the main text, disrupting sentence flow and introducing noise. Complex layouts commonly found in UGTEs, including interleaved text and images, multi-level lists, and nested tables or figures, further degrade the coherence of the extracted content. Consequently, the raw text is often disordered and unsuitable for direct use by language models. Each collected PDF thus required pre-processing to reconstruct a clean, semantically consistent, and logically structured plain-text representation.

\begin{figure*}[thpb]
    \centering
    \includegraphics[width=1\linewidth]{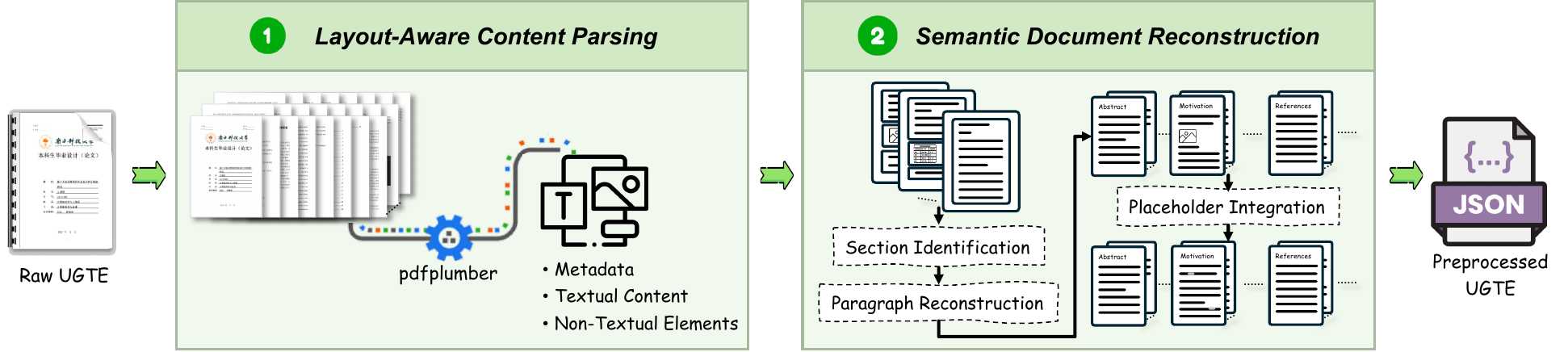}
    \caption{Overview of the UGTE data pre-processing pipeline. Consisting of two sequential stages, \ie, layout-aware content parsing and semantic documentat reconstruction, it transforms PDF-formatted theses into structured, JSON-based plain-text representations optimized for processing by large language models.
    }
    \label{fig:pre-processing}
\end{figure*}

To address these issues, a pre-processing pipeline is developed, as illustrated in Fig.~\ref{fig:pre-processing}. It converts PDF-formatted theses into plain-text representations compatible with LLMs. The overall process includes the following steps.

\noindent \textbf{Layout-Aware Content Parsing}: We use \textbf{\textit{pdfplumber}} to extract content from PDF files. The extracted elements include textual content such as titles, paragraphs, and footnotes; non-textual content such as tables and figures; and layout metadata such as font size, font weight, and absolute position. These features support reliable Differentiation between content types and structural roles throughout the document. To ensure compatibility with LLMs that process text only, each non-textual element is replaced by a semantic placeholder that retains its original caption. This preserves document structure and reference integrity in the absence of visual input.

\noindent \textbf{Semantic Document Reconstruction}: The raw output of layout-aware parsing often appears fragmented or disordered due to the absolute positioning in PDFs. To restore a coherent structure, section boundaries are first identified using common titles such as `Abstract' and `References', along with numeric patterns. The fragmented lines are then merged into complete and meaningful paragraphs based on spacing, indentation, and punctuation cues. Semantic placeholders for figures and tables are inserted into appropriate locations based on their coordinates and nearby context. This step yields a document that preserves both semantic clarity and structural consistency for subsequent processing.

\noindent \textbf{Structured JSON Output}:
The final output consists of the processed text and associated metadata, including thesis title, section labels, figure, and table references. All information is serialized in a structured JSON format.

Through this multi-stage pipeline, each thesis is transformed into a semantically coherent and structurally faithful format suitable for LLM-based evaluation. 

\section{Experimental Results}
\subsection{Experimental Settings}

\subsubsection{LLMs for Evaluation}
To assess PEMUTA's effectiveness and generalizability under realistic conditions, we evaluate it on diverse state-of-the-art LLMs via public APIs. All models are employed in a training-free setting, without parameter tuning or model-specific adaptation. This setup aligns with practical deployment scenarios in educational environments. As multi-granular UGTE assessment requires document-level understanding, sufficient context length is the key prerequisite. Models unable to accommodate full-length UGTEs, which often exceed $8,000$ – $10,000$ tokens, are hence excluded from consideration, leading to the following  models for evaluation.  

\noindent \textbf{Qwen-series}: Developed by Alibaba, the Qwen series comprises a family of Transformer-based LLMs optimized for NLP tasks. From this series, we select three representative models: \textit{Qwen-Plus}~\cite{yang_2024_qwenplus}, \textit{Qwen2.5-14B-1M}~\cite{yang_2025_qwen25-1MTechnicalReport}, and \textit{Qwen-Turbo-Latest}~\cite{yang_2025_qwen25-1MTechnicalReport}, all of which are well-suited for long-document processing and thus applicable to UGTE assessment. 
  \textit{Qwen-Plus}, with a maximum context window of 131K tokens, is officially recommended for its balanced trade-off among output quality, inference speed, and cost-efficiency. 
  \textit{Qwen2.5-14B-1M} and \textit{Qwen-Turbo-Latest} extend the context window to 1 million
  Both \textit{Qwen-Plus} and \textit{Qwen-Turbo-Latest} adopt a Mixture-of-Experts (MoE) architecture to support faster inference and reduced computational cost. They exhibit strong capabilities in instruction following, commonsense reasoning, and multilingual understanding. According to official documentation, \textit{Qwen-Plus} generally yields higher inference quality, while \textit{Qwen-Turbo-Latest} provides faster response.

  \noindent \textbf{DeepSeek-series}:
  Released by DeepSeek AI, this series emphasizes cost-efficiency and robust performance in complex reasoning tasks. Two representative models, \ie, \textit{DeepSeek-V3}~\cite{deepseekai_2024_deepseekv3} and \textit{DeepSeek-R1}~\cite{DeepSeek-AI_2025_DeepSeekR1} are employed, both adopting MoE architectures and supporting context lengths of 128K and 1M tokens, respectively, sufficient for processing full-length UGTEs. 
  \textit{DeepSeek-V3} serves as a general-purpose model, excelling in writing, summarization and multilingual translation, well-suited for holistic UGTE quality assessment. 
  \textit{DeepSeek-R1}, on the other hand, is specifically engineered for reasoning-intensive tasks, incorporating chain-of-thought (CoT) capabilities and self-verification mechanisms. These features enable step-by-step logical reasoning, making it particularly effective for supporting multi-granular UGTE assessment.

\subsubsection{Evaluation Metrics} 
Three commonly used evaluation metrics from automated essay scoring research~\cite{Alikaniotis2016AES, Song_2024_AESAER} are employed to assess the performance of the proposed framework. Let \(N\) denote the number of UGTEs.

\noindent \textbf{Mean Absolute Error (MAE)} computes the average absolute difference between predicted scores \( \hat{y}_i \) and ground-true scores \( y_i \):
\begin{equation}
\text{MAE} = \frac{1}{N} \sum_{i=1}^{N} \left| y_i - \hat{y}_i \right|,
\end{equation}

\noindent \textbf{Mean Squared Error (MSE)} penalizes larger errors more heavily and is defined as:
\begin{equation}
\text{MSE} = \frac{1}{N} \sum_{i=1}^{N} \left( y_i - \hat{y}_i \right)^2, 
\end{equation}

\noindent \textbf{Pearson Correlation Coefficient (PCC)} measures the linear correlation between predicted and expert-assigned scores:
\begin{equation}
\text{PCC} = \frac{\sum_{i=1}^{N} (y_i - \bar{y}) (\hat{y}_i - \bar{\hat{y}})}{\sqrt{\sum_{i=1}^{N} (y_i - \bar{y})^2} \cdot \sqrt{\sum_{i=1}^{N} (\hat{y}_i - \bar{\hat{y}})^2}}, 
\end{equation}
where \( \bar{y} \) and \( \bar{\hat{y}} \) are the means of the ground-truth and predicted scores, respectively.

MAE and MSE assess prediction accuracy, with lower values indicating better performance. PCC, on the other hand, reflects the degree of alignment between the model's performance in UGTE assessment and human expert evaluations, with higher values indicating stronger agreement. All three metrics are jointly used to provide a comprehensive view of both accuracy and robustness. Although Quadratic Weighted Kappa (QWK) is commonly used in traditional AES tasks, it is excluded in this task as it is tailored to discrete ordinal scales, whereas our scoring involves continuous-valued outputs.
  
\subsubsection{Implementation Details}

All model inferences are conducted via public APIs on Alibaba Cloud's Bailian platform using their official SDKs, without any fine-tuning or parameter-specific customization. This ensures that PEMUTA operates in a training-free, plug-and-play manner. All API requests are executed using the platform's default hyperparameters. To ensure output stability and avoid potential rate-limit violations, a delay of 30 seconds is applied between successive requests. For the few-shot prompting strategy, unless otherwise specified, \eg, in few-shot ablation analysis, the number of exemplars is empirically fixed at two. Three representative UGTE samples are selected from the MUTA dataset and manually constructed as the exemplar pool. For each inference, two exemplars are randomly sampled from this pool. To prevent data leakage and ensure fair evaluation, these samples are excluded from the final metric computation. In role-play prompting, the instruction is set to: \textit{`You are a university professor responsible for evaluating students' submitted undergraduate thesis.'}. 

\begin{figure}[!t]
    \centering
    \includegraphics[width=0.9\linewidth]{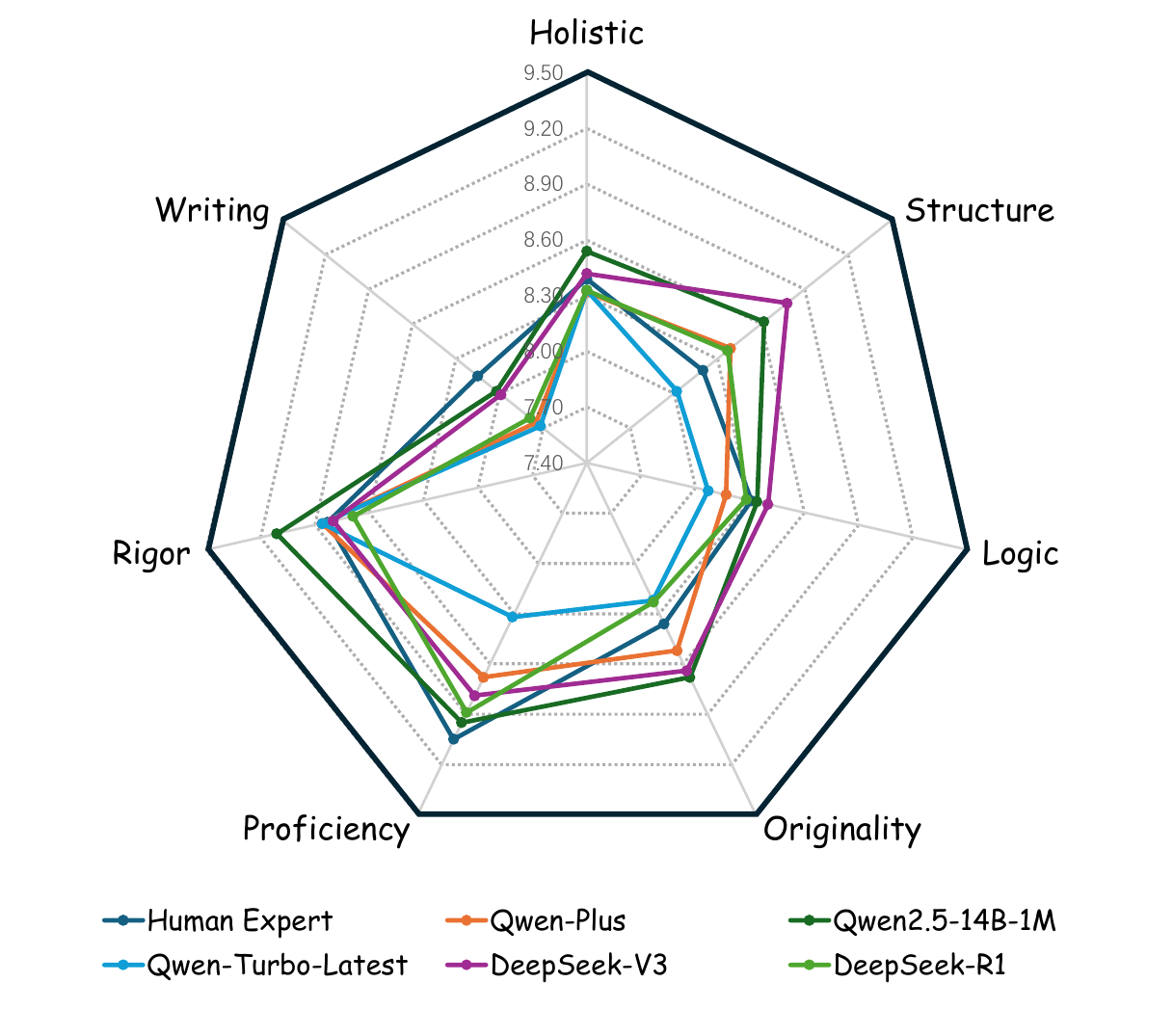}
    \caption{Radar chart summarizing holistic and dimension-specific mean scores for each model, with human expert annotations included for comparison.}
    \label{fig:MGResults}
\end{figure}
\subsection{Comparison with Standard Prompting Across LLMs}
\begin{table*}[htpb]
\caption{Comparison of the standard prompting strategy and the proposed PEMUTA framework across five LLMs on the MUTA dataset. Human expert scores are shown for reference. Best holistic metrics per model are in \textbf{bold}.}
\label{tab:fullresults}
\centering
\resizebox{\textwidth}{!}{
\footnotesize
\begin{tabular}{llcccccccccc}
\toprule%
\multirow{2.5}{*}{\textbf{Model}} & \multirow{2.5}{*}{\textbf{Prompt}} & \multicolumn{4}{c}{\textbf{Holistic}}
& \textbf{Structure} & \textbf{Logic} & \textbf{Originality} & \textbf{Writing} & \textbf{Proficiency} & \textbf{Rigor}\\
\cmidrule(r){3-6} \cmidrule(r){7-12} 
& & \textbf{MEAN} & \textbf{MAE $\downarrow$} & \textbf{MSE $\downarrow$} & \textbf{PCC $\uparrow$} & \textbf{MEAN} & \textbf{MEAN} & \textbf{MEAN} & \textbf{MEAN} & \textbf{MEAN} & \textbf{MEAN} \\
\midrule
\rowcolor{gray!30} \multicolumn{2}{c}{\textbf{Human Expert}} & \(8.39^{\pm.43}\) & -- & -- & -- & 8.20 & 8.31 & 8.36 & 8.15 & 9.05 & 8.83 \\
\midrule
    \multirow{2}{*}{\textbf{Qwen-Plus}}  & Standard & \(8.09^{\pm.55}\) & 0.44 & 0.45 & 0.25 & -- & -- & -- & -- & -- & -- \\
                                  & PEMUTA & \(8.32^{\pm.36}\) & \textbf{0.33} & \textbf{0.18} & \textbf{0.46} & 8.39 & 8.17 & 8.52 & 7.75 & 8.68 & 8.86 \\
    \midrule
    \multirow{2}{*}{\textbf{Qwen2.5-14B-1M}} & Standard & \(7.92^{\pm.60}\)& 0.59& 0.63& 0.25 & -- & -- & -- & -- & -- & -- \\
                                      & PEMUTA & \(8.54^{\pm.19}\)& \textbf{0.32} & \textbf{0.18 }& \textbf{0.39 }& 8.62 & 8.34 & 8.68 & 8.02 & 8.95 & 9.11\\
    \midrule
    \multirow{2}{*}{\textbf{Qwen-Turbo-Latest}}  & Standard & \(7.54^{\pm.71}\)& 0.89& 1.30& 0.14 & -- & -- & -- & -- & -- & -- \\
                                          & PEMUTA & \(8.33^{\pm.27}\)& \textbf{0.32}& \textbf{0.17}& \textbf{0.40} & 8.02& 8.07& 8.22&7.72 &8.32 &8.86\\
    \midrule
    \multirow{2}{*}{\textbf{DeepSeek-V3}} & Standard & \(7.99^{\pm.49}\)& 0.52 & 0.50 & 0.18 & -- & -- & -- & -- & -- & -- \\
                                   & PEMUTA & \(8.42^{\pm.25}\)& \textbf{0.34} & \textbf{0.19} & \textbf{0.29} & 8.78 & 8.40 & 8.64 & 7.99 & 8.79 & 8.80\\
    \midrule
    \multirow{2}{*}{\textbf{DeepSeek-R1}} & Standard & \(8.75^{\pm.47}\)& 0.57 & 0.42 & 0.29 & -- & -- & -- & -- & -- & -- \\
                                   & PEMUTA & \(8.33^{\pm.31}\)& \textbf{0.33} & \textbf{0.17} & \textbf{0.44} & 8.37 & 8.28 & 8.23 & 7.79 & 8.89 & 8.69\\
\bottomrule
\end{tabular}
}
\end{table*}

To evaluate the effectiveness of the proposed PEMUTA framework, we compare it against a standard prompting strategy across multiple LLMs on the MUTA dataset. The standard prompt follows conventional practice in prior LLM-based assessment research, using the instruction: \textit{`Holistically assess the given thesis on a 1–10 scale'}, applied to the same UGTE input across all models. Table~\ref{tab:fullresults} summarizes holistic-level results (mean score, MAE, MSE, PCC) as well as mean scores across the six SLOWPR dimensions.

Fig.~\ref{fig:MGResults} provides a visual summary of mean scores across all levels of granularity for all models. We have the following observations. 
\textbf{First}, across all evaluated models, PEMUTA consistently outperforms the standard prompt in holistic assessment. It achieves MAE reductions of $0.27$ – $0.57$, MSE decreases of $0.26$ – $1.13$, and PCC improvements of $0.14$ – $0.26$, indicating stronger alignment with expert ratings. Moreover, PEMUTA produces lower standard deviations in holistic scores while maintaining these improvements, suggesting better stability and consistency in UGTE assessment than the standard prompt. For example, on Qwen2.5-14B-1M, MAE decreases from $0.59$ to $0.32$, MSE from $0.63$ to $0.18$, PCC increases from $0.25$ to $0.39$, and the standard deviation decreases from $0.60$ to $0.19$.
\textbf{Second}, PEMUTA produces both holistic and dimension-specific scores, in contrast to the standard prompt, which yields only an overall rating. PEMUTA encourages multi-granular assessment, helping LLMs capture nuanced aspects of thesis quality and produce more interpretable and diagnostic evaluations. For instance, scores on \textit{Writing} are generally lower than those on \textit{Proficiency}, suggesting that students demonstrate stronger mastery of disciplinary knowledge while exhibiting relatively weak academic writing.
\textbf{Third}, scores vary across models and evaluation dimensions, suggesting model-specific tendencies. For instance, Qwen2.5-14B-1M assigns relatively higher scores on \textit{Writing} and \textit{Rigor}, with mean scores of $8.02$ and $9.11$, respectively, compared to other models whose scores generally range from $7.7$ to $8.0$ and $8.7$ to $8.9$. Meanwhile, Qwen-Turbo-Latest yields a noticeably lower average score of $8.32$ in \textit{Proficiency} compared to $8.68 \sim 8.95$ for other models.

\renewcommand{\arraystretch}{1.3}{
\begin{table}[t]
  \centering
\caption{Ablation study of major components on the MUTA dataset across models. For brevity, `H-P', `R-P', and `F-S' denote hierarchical prompting guidance, role-playing instruction, and few-shot exemplars. `Qwen2.5-14B' and `Qwen-Turbo' refer to `Qwen2.5-14B-1M' and `Qwen-Turbo-Latest', respectively. Best results are highlighted in \textbf{bold}.}
\resizebox{\linewidth}{!}{
\footnotesize
   \begin{tabular}{l|ccccccc}
    \toprule
   &    \textbf{H-G}&\textbf{R-P}&\textbf{F-S}&\textbf{MEAN}& \textbf{MAE $\downarrow$} & \textbf{MSE $\downarrow$} & \textbf{PCC $\uparrow$}\\
    \midrule  
    \multirow{4}{*}{\rotatebox{90}{\textbf{Qwen-Plus}}}  
        &\ding{55}&\ding{55}&\ding{55}& \(8.09^{\pm.55}\)& 0.44& 0.45&0.25\\
        &\ding{51}&\ding{51}&\ding{55}& \(8.02^{\pm.47}\)& 0.45& 0.38&0.34\\
        &\ding{51}&\ding{55}&\ding{51}& \(8.27^{\pm.29}\)& 0.34& 0.19&0.39\\
        &\ding{51}&\ding{51}&\ding{51}& \(8.32^{\pm.36}\)& \textbf{0.33}& \textbf{0.18}&\textbf{0.46}\\
    \midrule
    \multirow{4}{*}{\rotatebox{90}{\textbf{Qwen2.5-14B}}} 
        &\ding{55}&\ding{55}&\ding{55}& \(7.92^{\pm.60}\)& 0.59& 0.63&0.25\\
        &\ding{51}&\ding{51}&\ding{55}& \(8.36^{\pm.41}\)& 0.39& 0.24&0.32\\
        &\ding{51}&\ding{55}&\ding{51}& \(8.53^{\pm.13}\)& 0.34& 0.20& 0.21\\
        &\ding{51}&\ding{51}&\ding{51}& \(8.54^{\pm.19}\)& \textbf{0.32}&\textbf{0.18}& \textbf{0.39}\\
    \midrule
    \makecell{\multirow{4}{*}{\rotatebox{90}{\textbf{Qwen-Turbo}}}} 
        &\ding{55}&\ding{55}&\ding{55}& \(7.54^{\pm.71}\)& 0.89& 1.30& 0.14\\
        &\ding{51}&\ding{51}&\ding{55}& \(8.23^{\pm.42}\)& 0.47& 0.29& 0.28\\
        &\ding{51}&\ding{55}&\ding{51}& \(8.22^{\pm.38}\)& 0.34& 0.28& 0.24\\
        &\ding{51}&\ding{51}&\ding{51}& \(8.33^{\pm.27}\)& \textbf{0.32}& \textbf{0.17}& \textbf{0.40}\\
    \midrule
    \multirow{4}{*}{\rotatebox{90}{\textbf{DeepSeek-V3}}}  
        &\ding{55}&\ding{55}&\ding{55}& \(7.99^{\pm.49}\)& 0.52& 0.50&0.18\\
        &\ding{51}&\ding{51}&\ding{55}& \(8.44^{\pm.35}\)& 0.37& 0.25&0.21\\
        &\ding{51}&\ding{55}&\ding{51}& \(8.43^{\pm.25}\)& \textbf{0.34}& \textbf{0.19}&0.28\\
        &\ding{51}&\ding{51}&\ding{51}& \(8.42^{\pm.25}\)& \textbf{0.34}& \textbf{0.19}&\textbf{0.29}\\
    \midrule
    \multirow{4}{*}{\rotatebox{90}{\textbf{DeepSeek-R1}}}  
        &\ding{55}&\ding{55}&\ding{55}& \(8.75^{\pm.47}\)& 0.57& 0.42& 0.29\\
        &\ding{51}&\ding{51}&\ding{55}& \(8.68^{\pm.42}\)& 0.47& 0.33& 0.34\\
        &\ding{51}&\ding{55}&\ding{51}& \(8.17^{\pm.34}\)& 0.34& 0.18& 0.34\\
        &\ding{51}&\ding{51}&\ding{51}& \(8.33^{\pm.31}\)& \textbf{0.33}& \textbf{0.17}& \textbf{0.44}\\
    \bottomrule
    \end{tabular}%
  }
  \label{tab:componentAnalysis}%
\end{table}%
}

\subsection{Ablation Studies}
\subsubsection{Ablation of Major Components}
An ablation study is carried out on the MUTA dataset to evaluate PEMUTA's core components: Hierarchical prompting guidance (H-P), Role-Playing instruction (R-P), and Few-Shot exemplars (F-S). The comparison results are presented in Table~\ref{tab:componentAnalysis}. The holistic-level mean/std score, MAE, MSE, and PCC are employed for the performance comparison.

\begin{table*}[t]
  \centering
  \footnotesize
  \caption{Impact of the number of few-shot exemplars on PEMUTA's holistic performance. As the number of exemplars increases, the effectiveness of the proposed method in leveraging exemplar-based inductive reasoning becomes more pronounced. Best and second-best results across shot settings are marked in \textbf{bold} and \underline{underlined}, respectively.
  }    
  
    \begin{tabular}{lcccccccccccc}
    \toprule
    \multirow{2.5}{*}{\textbf{Model}} & \multicolumn{3}{c}{\textbf{0-shot}} &\multicolumn{3}{c}{\textbf{1-shot}} & \multicolumn{3}{c}{\textbf{2-shot}}& \multicolumn{3}{c}{\textbf{3-shot}} \\
    
    \cmidrule(r){2-4} \cmidrule(r){5-7} \cmidrule(r){8-10} \cmidrule(r){11-13}  & \textbf{MAE $\downarrow$} & \textbf{MSE $\downarrow$} & \textbf{PCC $\uparrow$} & \textbf{MAE $\downarrow$} & \textbf{MSE $\downarrow$} & \textbf{PCC $\uparrow$} & \textbf{MAE $\downarrow$} & \textbf{MSE $\downarrow$} & \textbf{PCC $\uparrow$} & \textbf{MAE $\downarrow$} & \textbf{MSE $\downarrow$} & \textbf{PCC $\uparrow$}  \\
    \midrule
    \textbf{Qwen-Plus} &  0.45 & 0.38 & 0.34 & \underline{0.36} & \underline{0.23} & \underline{0.37} &  \textbf{0.33} & \textbf{0.18} & \textbf{0.46} & 0.41 & 0.28 & 0.25 \\
    \textbf{Qwen2.5-14B-1M} & 0.39 & 0.24 & 0.32 & \underline{0.34} & 0.21 & 0.32 & \textbf{0.32} & \textbf{0.18} & \underline{0.39} & 0.36 & \underline{0.19} & \textbf{0.61}\\
    \textbf{Qwen-Turbo-Latest} & 0.47 & 0.29 & 0.28 & \underline{0.34} & \underline{0.22} & 0.30 & \textbf{0.32}&\textbf{0.17}&\underline{0.40}& 0.41 & 0.31 & \textbf{0.47}\\
    \textbf{DeepSeek-V3} & 0.37 & 0.25 & 0.21 & \underline{0.33} & \textbf{0.19} & 0.24 & \textbf{0.34} & \textbf{0.19} & \underline{0.29} & 0.39& 0.24& \textbf{0.42}\\
    \textbf{DeepSeek-R1} & 0.47 & 0.33 & 0.34  & \textbf{0.33} & \textbf{0.17} &  \textbf{0.49} & \textbf{0.33} & \textbf{0.17} & \underline{0.44} & \underline{0.41} & \underline{0.24} & 0.42 \\
    \bottomrule
    \end{tabular}%
  \label{tab:ablation_fewshot}%
\end{table*}%

As shown in Table~\ref{tab:componentAnalysis}, the variant lacking all three components corresponds exactly to the standard prompt strategy and performs the worst across all models. In contrast, the full PEMUTA configuration, incorporating all three components, consistently outperforms the other variants, achieving the lowest error rates and highest correlation with human scores. Furthermore, removing either the Role-Playing instruction (R-P) or the Few-Shot exemplars (F-S) results in a measurable performance decline, demonstrating that both components contribute meaningfully and synergistically to aligning the assessment with human expert judgment.

\subsubsection{Ablation of Few-shot Prompting}
The PEMUTA framework incorporates few-shot prompting to provide exemplar-based demonstrations for multi-granular UGTE assessment. Intuitively, these examples help guide the model toward human-like evaluative reasoning by exposing it to domain-specific assessment patterns. To empirically validate this design, we vary the number of in-context exemplars from 0 to 3 and evaluate the holistic-level performance using MAE, MSE, and PCC. The results are summarized in Table~\ref{tab:ablation_fewshot}.

As shown in Table~\ref{tab:ablation_fewshot}, the performance of PEMUTA exhibits a consistent upward trend as the number of exemplars increases from $0$ to $2$ across all models. Compared to the 0-shot setting, the 2-shot configuration reduces MAE by approximately $0.03$ - $0.15$, decreases MSE by about $0.06$ - $0.20$, and improves PCC by roughly $0.07$ - $0.12$. For instance, on Qwen-Turbo-Latest, MAE drops from $0.47$ (0-shot) to $0.32$ (2-shot), while PCC improves from $0.28$ to $0.40$. These results suggest that exemplar-based prompting effectively enhances model alignment with expert ratings. However, the trend does not persist beyond two exemplars. In several cases, such as Qwen-Plus and DeepSeek-R1, increasing the shot count to three leads to degraded performance metrics. This decline is potentially due to the prompt length approaching or exceeding the model's effective context window, or introducing additional noise that distracts the model's reasoning. This finding highlights a practical trade-off between adding more exemplars to enrich context while managing prompt length to avoid negatively impacting model reasoning.

\subsection{Qualitative Analysis}
\label{sec:qualitative}

\begin{figure*}[thpb]
    \centering
    \includegraphics[width=\linewidth]{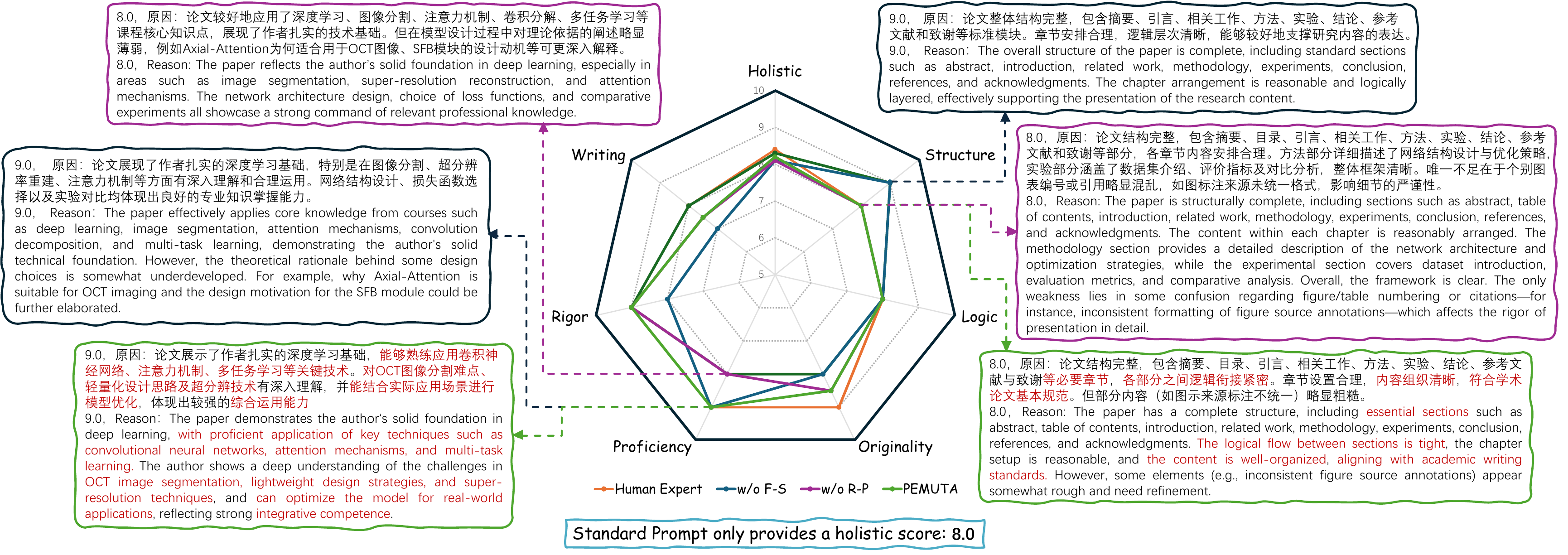}
    \caption{Case study of multi-granular assessment for a sample from the MUTA dataset by Qwen-Plus under four prompting configurations: Standard Prompt, full PEMUTA, and two ablated PEMUTA variants. The radar plot illustrates model-assigned scores across six fine-grained dimensions and one holistic dimension for each multi-granular-capable configuration, alongside human expert ratings. The Standard Prompt, which only outputs a holistic score, is shown separately. Representative score justifications are provided for the \textit{Structure} and \textit{Proficiency} dimensions. Excerpts uniquely generated or better articulated by full PEMUTA are highlighted in \red{red}.} 
\label{fig:qualitative}
\end{figure*}

To intuitively assess the proposed PEMUTA in guiding LLMs to perform pedagogically meaningful multi-granular UGTE assessment, we present a case study in Fig.~\ref{fig:qualitative}. This example is taken from the MUTA dataset and evaluated by Qwen-Plus under four prompting configurations: (1) Standard Prompt, (2) PEMUTA without few-shot prompting, (3) PEMUTA without role-play prompting, and (4) Full PEMUTA. 

As expected, the standard prompt yields only a holistic score, while PEMUTA and its ablated variants support multi-granular assessment. The radar chart shows the holistic rating and model-predicted scores across the six SLOWPR dimensions, \ie, \textit{Structure}, \textit{Logic}, \textit{Originality}, \textit{Writing}, \textit{Proficiency}, and \textit{Rigor}. All three multi-granular-capable setups generate score justifications that are consistently relevant to the designated dimensions, indicating that PEMUTA-style prompting effectively activates latent domain knowledge within LLMs to support dimension-aware evaluation. Among the three, full PEMUTA aligns most closely with expert judgments, particularly in dimensions such as \textit{Structure} and \textit{Proficiency}, suggesting that the few-shot and role-playing components act synergistically to enhance pedagogical sensitivity and promote fine-grained reasoning.

Representative score justifications further demonstrate the advantage of full PEMUTA. In \textit{Structure}, it generates more specific and pedagogically grounded feedback, referencing segments such as `essential sections,' `tight logical flow,' and `organization consistent with academic writing standards.' In contrast, the ablated variants tend to offer vague or superficial comments. In \textit{Proficiency}, full PEMUTA more accurately diagnoses students’ methodological competence by referencing techniques such as `convolutional neural networks', `attention mechanisms', and `multi-task learning', while also connecting them to evaluative criteria in UGTE, such as `optimizing models for real-world applications' and `reflecting strong integrative competence'.

In summary, this case study illustrates that the proposed full PEMUTA enables pedagogically meaningful multi-granular assessment by producing scores that closely align with expert ratings, while also delivering informative and specific feedback that reflects core evaluation criteria in UGTE assessment.

\section{Conclusion and Future Work}

In this paper, a pedagogically-enriched multi-granular UGTE assessment framework, PEMUTA, is proposed to effectively activate domain-specific knowledge from large language models for undergraduate thesis evaluation. PEMUTA incorporates hierarchical two-step prompting grounded in Vygotsky's theory and Bloom's Taxonomy, enabling structured assessment across six fine-grained dimensions: Structure, Logic, Originality, Writing, Proficiency, and Rigor (SLOWPR), followed by holistic synthesis. This hierarchical design effectively mitigates LLM attention distraction by guiding focused evaluation along each dimension, thereby leveraging embedded academic knowledge more precisely to generate assessments that reflect both students' academic performance and their developmental potential. Two in-context learning techniques, namely, few-shot prompting and role-play prompting, are further jointly utilized to maximize pedagogical alignment, enhancing consistency with expert judgments without additional fine-tuning. Extensive experiments demonstrate that PEMUTA consistently outperforms standard holistic prompting strategies, achieving lower MAE and MSE as well as higher PCC with expert ratings, and addressing the limitation of existing methods that cannot produce fine-grained, dimension-specific assessments. 

In future, we will extend PEMUTA into multimodal assessment frameworks that incorporate code artifacts, presentation recordings, and other relevant learning outputs, thereby enabling more comprehensive evaluation of students' disciplinary competencies and practical skills.

\section*{Acknowledgments}
This work was supported in part by the Guangdong Higher Education Teaching Reform Project under Grant SJZLGC202202; in part by the Southern University of Science and Technology Teaching Reform Project under Grant XJZLGC202414; and in part by the SUSTech Research Key Project under Grant SUST23Z01.

\bibliographystyle{IEEEtran}
\bibliography{mybibfile}

\newpage
\vfill

\end{document}